# Corrections to Kusnetz Method for Measurements of Radon Progeny Concentrations in Air

**Omar Nusair**

*Department of Physics and Astronomy, University of Alabama, Tuscaloosa 35487, U.S.A.*

*E-mail:* oynusair@ua.edu

ABSTRACT: In 1956, H.L. Kusnetz proposed a quick method for radon progeny concentration measurement in mine atmosphere using a single gross-alpha count of a membrane-filtered air. The method is still widely used today and is based on a number of impractical assumptions. An instantaneous sampling time (less than ten seconds), is one of these assumptions that ignores the build-up and decay of the progeny on the filter paper during the sampling period. Of special concern is the $^{214}$Bi decay during the sampling period, since $^{214}$Po's alphas are lost during the sampling time and cannot be accounted for during the counting time. In addition, the method assumes that $^{214}$Bi activity during the counting period is constant. This inaccurate assumption can result in a smaller count rate, especially when counting times are long. Hence underestimated working levels are expected when using Kusnetz factors without correcting for the sampling and counting times. In this technical report, exact sampling and counting time corrections to the method are provided along with the updated Kusnetz factors that correspond to common equilibrium conditions to correctly estimate the Working Level in air. Additionally, time corrections to the commonly used self-absorption correction formula and the lower level of detectability (LLD) equation used for any sample measurement are given.

# Contents

## 1. Introduction

Radon-222 and its short-lived decay products, called progeny, are some of the airborne radionuclides of concern at uranium mines, mills, and In-Situ Recovery (ISR) facilities. It is a decay product of the naturally occurring $^{238}$U present in soil, with the outdoor radon concentrations typically around 0.4 pCi/L (14.8 Bq/m$^3$) while the indoor concentrations are 1.3 pCi/L (37 Bq/m$^3$) [1]. Therefore, radon public dose is a concern at facilities that produce yellow cake or generally process raw uranium ore. Hence, uranium mines, mills, and ISR facilities are required by the regulatory authorities to monitor their effluent releases of radon and its progeny in the atmosphere. In addition, individuals who work in these facilities are directly exposed to radon and its progeny from the ore, uranium extraction and processing stages, and from the mill tailings. Moreover, members of the general public at the nearest location to a uranium production facility are exposed to radon and its progeny.

A survey of the cumulative exposure to the radon progeny concentration in air can be made by conducting a measurement using a known volume of air. In this context, the Potential Alpha Energy Concentration (PAEC) quantity is widely used, which is a measure of the total energy of alpha particles that can potentially be emitted in one liter of air if radon progeny undergo complete decay to $^{210}$Pb. An important advantage of this quantity is that there is no need to specifically determine concentrations of each radon progeny, where the Working Level (WL) unit, introduced by [2], was assigned as a unit for the PAEC measurement. For example, a 100 pCi/L of the volumetric activity concentration of radon progeny, in equilibrium with radon, will result in one WL of PAEC. In the above example, the total decay of radon decay products will ultimately result in the emission of 129,194.7 MeV of alpha-particle energy per liter of air (i.e. one WL of PAEC). However, the use of PAEC and WL does not require $^{222}$Rn and its progeny to be in 100% equilibrium. In fact, radon progeny are quite often out of equilibrium with each other and with their precursor $^{222}$Rn. Consequently, and regardless of which short-lived progeny emits the alpha radiation, the potential alpha energy in a unit volume of air at a given point in time is what matters. Additionally, the closer the equilibrium factor to 100%, the greater the dose per unit radon concentration.

An environmental sampling program goal at a uranium mine, mill, or ISR facility must ensure that radon-related public dose rates associated with the operations of the facility are below regulatory limits. Therefore, the monitoring programs at these sites must be able to monitor low working levels of PAEC in public and occupational areas. Hence, Kusnetz Method became widely used for measurement of the radon progeny concentration in air at uranium mines, mills and ISR facilities, for its easy application and quick results. The following sections briefly summarize the method. Time corrections are provided in Section 4.3. A corrected self-absorption calculation method is provided in Section 5, and a formula for calculating a measurement Lower-Level of Detectability (LLD) is given in Section 6.

## 2. The Kusnetz Method

H. Kusnetz [3] proposed a quick method for radon progeny concentration measurement using a gross-alpha count of a filtered air on cellulose membrane or glass fiber filters. The air is typically sampled over three minutes. A waiting time is required for $^{218}$Po to completely decay away before counting starts, and to eliminate the dependency on the equilibrium condition as will be explained in Section 4.3. This waiting period, between 40 and 90 minutes, is long enough to ensure that $^{218}$Po's 6.003-MeV-alpha particles are not overlapping with the counting of the $^{214}$Po's alpha particles (E= 7.687 MeV) during the counting period. This is because the method is based on the use of a scintillator such as zin sulfide that has poor energy resolution for alpha radioactivity measurements. The count times typically range from one to ten minutes, then the radon progeny concentration in air can be determined using the following formula:

$$PAEC = \frac{R \cdot (1 + SA)}{K \cdot V \cdot \varepsilon} \qquad (2.1)$$

where, $R$ is the net (background subtracted) alpha count rate in (cpm), $K$ is Kusnetz time correction factor (dpm/L/WL) that corresponds to a given delay time, $V$ is the total volume of sampled air in liters (L), $\varepsilon$ is the alpha detection efficiency (cpm/dpm), and $SA$ is a correction factor to account for self-absorption of the alpha particles emitted from the progeny, the formula is provided in Section 5.

## 3. Impractical Hypotheses of the Kusnetz Method

The derivation of Kusnetz time factors assumes an instantaneous sampling time (less than ten seconds) [3]. This impractical assumption simply ignores the build-up and decay of the progeny on the filter paper during the sampling period. The decay of $^{214}$Bi during the sampling period cannot be accounted for during the alpha counting phase of the air filter. Moreover, since no correction is made, the method assumes that $^{214}$Bi activity during the counting period is constant, which results in a smaller count rate, and therefore, underestimated working levels are expected. As a result, Kusnetz factors must be updated to correct for the sampling and counting times.

Second, the use of Kusnetz factors without taking into considerations their quoted uncertainties in [3] assumes the presence of 100% equilibrium condition between radon progeny. In fact, all Kusnetz factors were derived based on a 1:1:1 equilibrium condition (i.e. concentration ratios of $^{218}$Po: $^{214}$Pb: $^{214}$Bi). However, in the 1956 paper Kusnetz estimated the errors associated with varying the equilibrium conditions to range from 1.2 to 12 (dpm/L/WL), which also depends on the delay time after sampling.

Last, the method is valid only when no $^{220}$Rn exists in the atmosphere. Radon-220 (also known as thoron) is a member of the $^{232}$Th decay chain. Nevertheless, when it is produced in the ore soil, $^{220}$Rn has shorter time to emanate due to its short half-life. Therefore, the assumption of no $^{220}$Rn being sampled on the filter may hold true even when $^{232}$Th and $^{238}$U may be present in the ore deposits with relatively same concentrations.

## 4. Sampling- and Counting-Time Corrections and the Updated Kusnetz Factors

As explained above the radon decay products are filtered out of the air, while radon itself is a noble gas and thus, it will not be collected on the filter. The collected $^{218}$Po is unstable, therefore some of it will already decay while still sampling the air. Two phases can be distinguished for the derivation of Kusnetz factors, namely the sampling or collection phase, and the delay and counting phase. In each phase, a Bateman's equation that describes the growth and decay of a particular daughter is solved and the solution is used for determining the next daughter concentration in the chain in one phase to be used for the next phase.

### 4.1 Sampling Phase

During the filtration process the activity of each individual progeny can be described by build-up and decay terms in the following Bateman's equations, which can be attributed to the derivations of the exact collection- and counting-time corrected Kusnetz factors:

$$\frac{dN_{Po}}{dt} = \dot{v} \cdot n_{Po} - \frac{N_{Po}}{\tau_{Po}} \qquad (4.1.1)$$

$$\frac{dN_{Pb}}{dt} = \dot{v} \cdot n_{Pb} + \frac{N_{Po}}{\tau_{Po}} - \frac{N_{Pb}}{\tau_{Pb}} \qquad (4.1.2)$$

$$\frac{dN_{Bi}}{dt} = \dot{v} \cdot n_{Bi} + \frac{N_{Pb}}{\tau_{Pb}} - \frac{N_{Bi}}{\tau_{Bi}} \qquad (4.1.3)$$

where, $N_{Po}$, $N_{Pb}$, and $N_{Bi}$ are unitless quantities which represent the number of filtered atoms of $^{218}$Po, $^{214}$Pb and $^{214}$Bi, respectively, $\dot{v}$ is the pumping rate in unit of (L/min), $n_{Po}$, $n_{Pb}$, and $n_{Bi}$ are the WL-normalized volumetric atom concentrations (atoms/L/WL) of $^{218}$Po, $^{214}$Pb and $^{214}$Bi, respectively, that result in one WL, and $\tau_{Po}$, $\tau_{Pb}$, and $\tau_{Bi}$ are the mean life times of $^{218}$Po (4.4695 min), $^{214}$Pb (30.0393 min), and $^{214}$Bi (28.7096 min), respectively.

If an equilibrium condition of 100% exists between all the progeny, then the corresponding values of $n_{Po}$, $n_{Pb}$, and $n_{Bi}$ are: 992.2, 8666.7, and 6373.5 (atoms/L/WL), respectively. In general, the WL-normalized volumetric atom concentrations can be calculated using the following general formula:

$$n_i = 222(\frac{dpm}{L}) \cdot q_i(\frac{\%}{WL}) \cdot \tau_i(min) \qquad (4.1.4)$$

where, $i$ represents the short-lived radon daughters $^{218}$Po, $^{214}$Pb, and $^{214}$Bi, and $q_i$ is the corresponding WL-normalized equilibrium factor for nuclide $i$.

For a sampling time of ($T_s$), the solutions for equations 4.1.1, 4.1.2 and 4.1.3 are given in terms of activities in equations 4.1.5, 4.1.6, and 4.1.7, respectively, as follows:

$$A_{Po}(T_s) = \dot{v} \cdot \alpha(T_s) \qquad (4.2.5)$$

$$A_{Pb}(T_s) = \dot{v} \cdot \beta(T_s) \qquad (4.3.6)$$

$$A_{Bi}(T_s) = \dot{v} \cdot \gamma(T_s) \qquad (4.4.7)$$

where, $\alpha(T_s)$, $\beta(T_s)$, and $\gamma(T_s)$ are symbolized sampling time factors associated with the $^{218}$Po, $^{214}$Pb and $^{214}$Bi build-ups and decays during the sampling phase, respectively, which are given in the following equations:

$$\alpha(T_s) = n_{Po} \cdot \left(1 - e^{-\frac{T_s}{\tau_{Po}}}\right) \qquad (4.5.8)$$

$$\beta(T_s) = \left[(n_{Po} + n_{Pb}) \cdot \left(1 - e^{-\frac{T_s}{\tau_{Pb}}}\right)\right] + \left[n_{Po} \cdot \frac{\tau_{Po}}{(\tau_{Po} - \tau_{Pb})} \cdot (e^{-\frac{T_s}{\tau_{Pb}}} - e^{-\frac{T_s}{\tau_{Po}}})\right] \qquad (4.6.9)$$

$$\gamma(T_s) = \Bigg\{ \left[(n_{Po} + n_{Pb} + n_{Bi}) \cdot \left(1 - e^{-\frac{T_s}{\tau_{Bi}}}\right)\right] \qquad (4.7.8)$$
$$+ \left[(\frac{n_{Pb}}{\tau_{Pb}} - \frac{n_{Po}}{\tau_{Po}} - \frac{n_{Pb}}{\tau_{Po}}) \cdot \frac{\tau_{Po} \cdot \tau_{Pb}^2}{(\tau_{Po} - \tau_{Pb}) \cdot (\tau_{Pb} - \tau_{Bi})} \cdot (e^{-\frac{T_s}{\tau_{Bi}}} - e^{-\frac{T_s}{\tau_{Pb}}})\right]$$
$$+ \left[n_{Po} \cdot \frac{\tau_{Po}^2}{(\tau_{Po} - \tau_{Pb}) \cdot (\tau_{Po} - \tau_{Bi})} \cdot (e^{-\frac{T_s}{\tau_{Bi}}} - e^{-\frac{T_s}{\tau_{Po}}})\right] \Bigg\}$$

## 4.2 Delay and Counting Phases

To derive Kusnetz factors, with the corrections for the sampling and counting times, one needs to develop a model that describes the activity of $^{214}$Po (or equivalently $^{214}$Bi), as a function of delay time, which is primarily targeted by the alpha counting in this method. In this phase of time there is no pumping of air, therefore the term $\dot{v}$ is set to zero, and only sequential decays are responsible for the decay and build-ups of the progeny according to the following equations:

$$\frac{dN_{Po}}{dt} = -\frac{N_{Po}}{\tau_{Po}} \qquad (4.2.1)$$

$$\frac{dN_{Pb}}{dt} = \frac{N_{Po}}{\tau_{Po}} - \frac{N_{Pb}}{\tau_{Pb}} \qquad (4.2.2)$$

$$\frac{dN_{Bi}}{dt} = \frac{N_{Pb}}{\tau_{Pb}} - \frac{N_{Bi}}{\tau_{Bi}} \qquad (4.2.3)$$

After solving equations 4.2.1, 4.2.2, and 4.2.3, the yield is the activity of $^{214}$Bi, $A_{Bi}(t)$, at any time ($t$) after end of sampling, which is given as follows:

$$A_{Bi}(t) = A_{Bi}(T_s) \cdot e^{-\frac{t}{\tau_{Bi}}} - A_{Pb}(T_s) \cdot \frac{\tau_{Pb}}{(\tau_{Pb} - \tau_{Bi})} \cdot \left(e^{-\frac{t}{\tau_{Bi}}} - e^{-\frac{t}{\tau_{Pb}}}\right) + A_{Po}(T_s) \cdot \frac{\tau_{Po}}{(\tau_{Po} - \tau_{Bi})} \quad (4.2.4)$$
$$\cdot \left[\frac{\tau_{Pb}}{(\tau_{Pb} - \tau_{Bi})} \cdot \left(e^{-\frac{t}{\tau_{Bi}}} - e^{-\frac{t}{\tau_{Pb}}}\right) - \frac{\tau_{Po}}{(\tau_{Po} - \tau_{Bi})} \cdot \left(e^{-\frac{t}{\tau_{Bi}}} - e^{-\frac{t}{\tau_{Po}}}\right)\right]$$

The contributions from the three activity components $A_{Bi}(T_s)$, $A_{Pb}(T_s)$, and $A_{Po}(T_s)$ in equation 4.2.4 are plotted as functions of delay time (i.e. time after end of sampling) in Figure [1], and the sum of the three functions is a plot of equation 4.2.4. The behavior of $A_{Bi}(t)$ is clearly dependent on the equilibrium condition that exists between radon progeny. It is also clear that the assumption of constant activity between $t = 40$ and $t = 90$ (min) is wrong, hence for counting intervals longer than one minute there needs to be a correction that considers the change of $^{214}$Bi population on the filter during counting time. Therefore, it is necessary to correct for that while deriving Kusnetz factors.

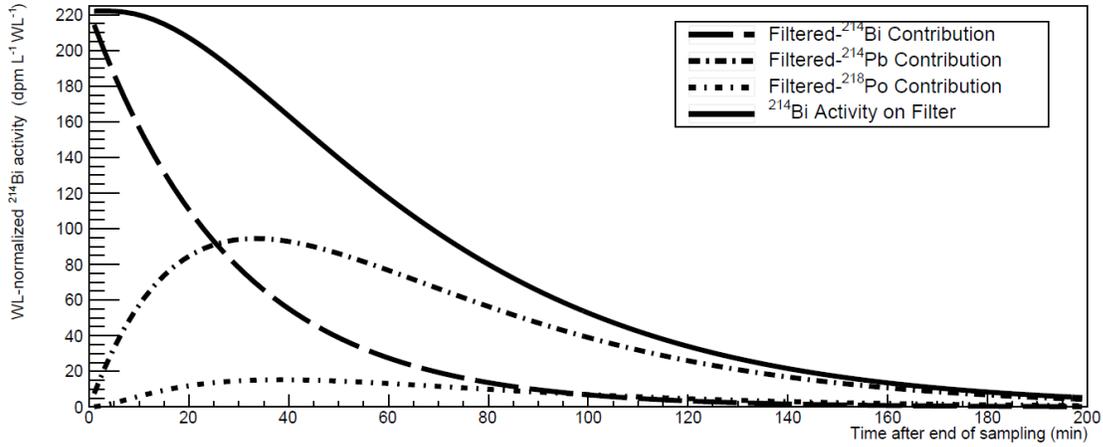

**Figure 1.** A theoretical activity curve of $^{214}$Bi (solid curve) after end of air sampling on a filter as a function of delay time. The dashed curves are for the individual contributions from $^{218}$Po, $^{214}$Pb, and $^{214}$Bi filtered from a hypothetically sampled air that contains 100 pCi/L of $^{222}$Rn in 100% equilibrium with all its short-lived progeny (i.e. 1 WL/L of radon-progeny concentration in air). This mixture of radon progeny results in one WL.

## 4.3 Corrected Kusnetz Factors

The Kusnetz factor in units of (dpm/L/WL) at any given delay time after end of sampling ($t$) is obtained by integrating the above equation between counting start time $t_1$ and counting stop time $t_2$ (both are determined from the end of sampling), and dividing by $\dot{v} \cdot T_s \cdot (t_2 - t_1)$ to normalize the number of resulted decays per unit volume ($Volume = \dot{v} \cdot T_s$) per unit counting time ($t_2 - t_1$) as follows:

$$K(t) = \frac{1}{T_s \cdot (t_2 - t_1)} \left\{ \gamma(T_s) \cdot \tau_{Bi} \cdot \left(e^{-\frac{t_1}{\tau_{Bi}}} - e^{-\frac{t_2}{\tau_{Bi}}}\right) - \beta(T_s) \cdot \frac{\tau_{Pb}}{\tau_{Pb} - \tau_{Bi}} \right. \quad (4.3.1)$$
$$\cdot \left[\tau_{Bi} \cdot \left(e^{-\frac{t_1}{\tau_{Bi}}} - e^{-\frac{t_2}{\tau_{Bi}}}\right) - \tau_{Pb} \cdot \left(e^{-\frac{t_1}{\tau_{Pb}}} - e^{-\frac{t_2}{\tau_{Pb}}}\right)\right] + \alpha(T_s) \cdot \frac{\tau_{Po}}{\tau_{Po} - \tau_{Pb}}$$
$$\cdot \left[\frac{\tau_{Pb}}{\tau_{Pb} - \tau_{Bi}} \cdot \langle \tau_{Bi} \cdot \left(e^{-\frac{t_1}{\tau_{Bi}}} - e^{-\frac{t_2}{\tau_{Bi}}}\right) - \tau_{Pb} \cdot \left(e^{-\frac{t_1}{\tau_{Pb}}} - e^{-\frac{t_2}{\tau_{Pb}}}\right)\rangle - \frac{\tau_{Po}}{\tau_{Po} - \tau_{Bi}} \right.$$
$$\left. \left. \cdot \langle \tau_{Bi} \cdot \left(e^{-\frac{t_1}{\tau_{Bi}}} - e^{-\frac{t_2}{\tau_{Bi}}}\right) - \tau_{Po} \cdot \left(e^{-\frac{t_1}{\tau_{Po}}} - e^{-\frac{t_2}{\tau_{Po}}}\right)\rangle\right]\right\}$$

There is a notable symmetry between the sampling and counting times such that when interchanging the two times one gets the exact same Kusnetz factor. Therefore, it is not necessary to report individual factors for each set of sampling and counting times, but one can generate factors relevant to the sum of both times (or sum time). Table [1] is an example of the updated Kusnetz factors for different atmospheric equilibrium conditions. Figure [2] below shows the original Kusnetz factors in comparison with the corrected factors from this work. It is notable that the cases of 20 and 40 minutes of the sum time are largely deviated from the original curve (solid line in Figure [2]). As a result, the use of the original factors without correcting for the sampling and counting times will result in an underestimation of the working level by a factor of 40% and 17% for the 40 and 20 (min) of sum times, respectively.

**Table 1.** Time-corrected Kusnetz factors for different equilibrium conditions at different sum times.

| delay time (min) | K (dpm/L/WL) | sum time sampling + counting (min) | | | | | | | | | | | |
|---|---|---|---|---|---|---|---|---|---|---|---|---|---|
| | | aged air (100:100:100) (pCi/L) | | | | indoor [4] (177:106:71) (pCi/L) | | | | outdoor [5] (125:97:97) (pCi/L) | | | |
| | | 4 | 6 | 10 | 20 | 4 | 6 | 10 | 20 | 4 | 6 | 10 | 20 |
| 40 | 150 | 154 | 152 | 147 | 135 | 152 | 151 | 147 | 137 | 152 | 150 | 145 | 134 |
| 45 | 140 | 142 | 140 | 135 | 124 | 143 | 141 | 137 | 127 | 141 | 139 | 134 | 123 |
| 50 | 130 | 131 | 128 | 124 | 114 | 133 | 131 | 127 | 117 | 130 | 127 | 123 | 113 |
| 55 | 120 | 120 | 118 | 113 | 104 | 123 | 121 | 117 | 108 | 119 | 117 | 113 | 103 |
| 60 | 110 | 109 | 107 | 103 | 94 | 113 | 111 | 108 | 99 | 109 | 107 | 103 | 94 |
| 65 | 100 | 100 | 98 | 94 | 85 | 104 | 102 | 98 | 90 | 99 | 97 | 94 | 85 |
| 70 | 90 | 90 | 89 | 85 | 77 | 95 | 93 | 90 | 82 | 90 | 88 | 85 | 77 |
| 75 | 83 | 82 | 80 | 77 | 70 | 87 | 85 | 82 | 74 | 82 | 80 | 77 | 70 |
| 80 | 75 | 74 | 73 | 70 | 63 | 79 | 77 | 74 | 67 | 74 | 72 | 69 | 63 |
| 85 | 68 | 67 | 65 | 63 | 57 | 71 | 70 | 67 | 61 | 67 | 65 | 63 | 57 |
| 90 | 60 | 60 | 59 | 57 | 51 | 65 | 63 | 61 | 55 | 60 | 59 | 56 | 51 |

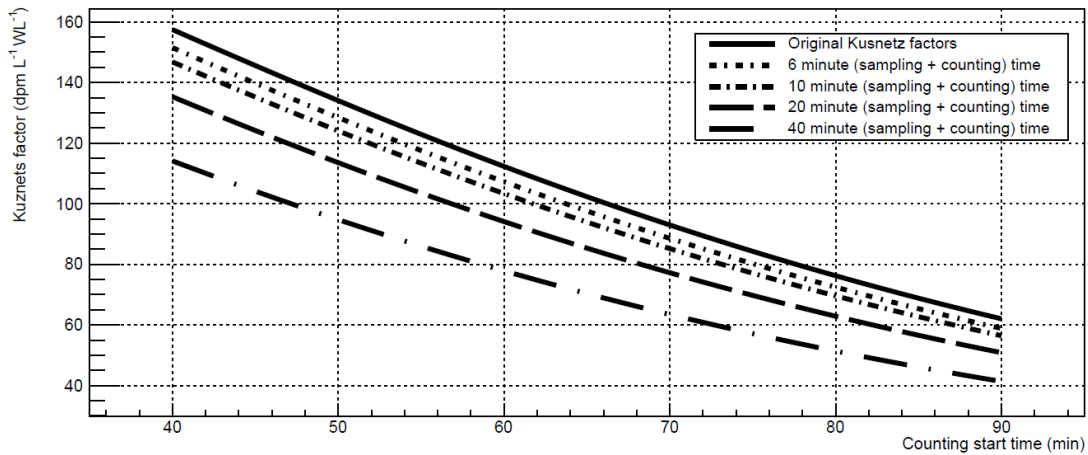

**Figure 2.** Original Kusnetz factors (solid curve) and different time-corrected factors (several dashed curves) from this work.

The relative equilibrium conditions between $^{218}$Po, $^{214}$Pb, and $^{214}$Bi, determine the behavior of the $^{214}$Bi activity curve. Especially in the early time, the decay behavior is largely impacted by these relative equilibrium concentrations of the progeny, as seen in Figure [3] below. The four examples shown in the figure are for different equilibrium ratios between radon and its progeny mixture (i.e. $q_{Po}$: $q_{Pb}$: $q_{Bi}$) that result in one WL for each case. The cases are: a 100% equilibrium case (1:1:1), a typical indoor equilibrium condition (1.77:0.60:0.71) [4], a typical outdoor equilibrium condition (1.25:0.97:0.97) [5], and an average equilibrium condition as measured in actual uranium mines (2.24:1.02:0.67) [6]. The ratios between the progeny mixture in each of the four cases are normalized to one WL.

From Figure [3], it is safe to ignore the equilibrium conditions when performing the Kusnetz Method since there is no significant difference is found between the four curves in the time window where the counting is performed.

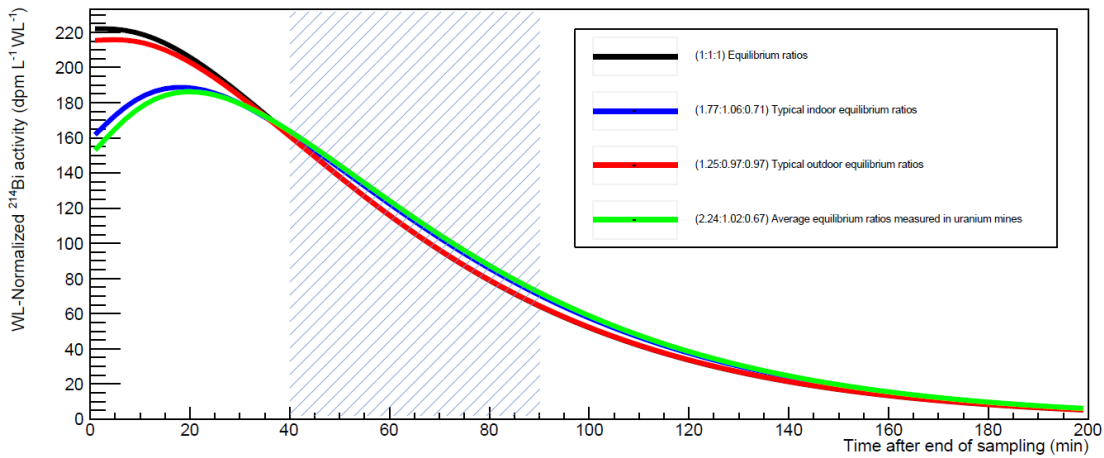

**Figure 3.** Different equilibrium conditions between $^{222}$Rn and its short-lived progeny result in different early behavior of $^{214}$Bi activity. The black curve is for a fully equilibrium condition in air that contains one WL of radon progeny. Typical indoor (blue), and outdoor (red) conditions, and the average progeny equilibrium condition as measured in actual uranium mines (green) are plotted for comparison. The shaded area is the applicability time window of Kusnetz Method. It is noted that no significant differences are observed in all the outlined cases of equilibrium in this comparison.

## 5. Self-Absorption Correction

Although most filters used in this method, e.g. cellulose filters and glass fiber and microfiber filters, are highly efficient in collecting radon progeny on the surface of the filter, it is assumed that some of the products might decay from inside the filter material. Also, collected dust and moisture content in the filter affect the transmission of alpha particles towards the detector. This results in an absorption of some of the alpha particles emitted towards the detector, depending on the effective thickness of the filter. To correct for this self-absorption, the following steps are followed in the uranium industry [7]: A sample of an air volume is collected on a filter of choice for five minutes, followed by 40 to 90 minutes of delay time. Then, the front face of the filter is counted for five minutes using the same detector used in the Kusnetz Method for working level measurement. The count rate in (cpm) is recorded as *A*. After that, the back face of the filter is counted for five minutes using the same detector under the same conditions, the count rate is recorded as *B* in (cpm). Then, an identical duplicate filter is used to cover the front face of the sample filter. Count rate with the duplicate inserted between the original filter and the detector, to act as an absorber, for five minutes should be recorded as *C* in (cpm). The correction factor, *SA*, used in equation 2.1 can be obtained using the following formula:

$$SA = \frac{B - C}{2A + B - C} \tag{5.1}$$

However, when determining the self-absorption factor for a given filter using long counting times (more than three minutes) the decay of the radon progeny could be significant. In fact, the $^{214}$Bi population in the first count of the front face of the filter is different than that when counting the back face, and different than that when counting the covered front face. This would cause the self-absorption factor to be biased low resulting eventually in an underestimated working level. Therefore, the corrections for sampling and counting time must be applied to the above formula, as follows:

$$SA = \frac{\frac{B}{K_B} - \frac{C}{K_C}}{\frac{2A}{K_A} + \frac{B}{K_B} - \frac{C}{K_C}} \tag{5.2}$$

where, $K_A$, $K_B$ and, $K_C$ are Kusnetz factors that correspond to the counting of the front side, backside, and the covered front side, respectively.

## 6. Lower Level of Detectability (LLD)

The LLD of a measured sample is the minimum level of radioactivity that can be potentially determined by a measurement process, such as the Kusnetz Method. It is important here to distinguish between the background coming from the daughters plating out on the filter surface due to poor storage conditions of the filter and that coming from the natural radon background (i.e. from unlicensed radon emissions). The latest should be determined in an established background location, such as an upwind monitoring station of a uranium facility, where only pure background is anticipated. For different counting intervals for the sample and the ambient background (including the plate-out background), the radon progeny detection LLD in (WL) for any sample measurement is calculated using the following formula [8], modified in this work to specify the suitable conversion factors:

$$LLD \ (WL) = \frac{3 + 3.29 \sqrt{R_B \cdot \frac{V_{sample}}{V_B} \cdot T_{sample} \cdot (1 + \frac{T_{sample}}{T_B})}}{K_B \cdot \varepsilon \cdot V_{sample} \cdot T_{sample}} \tag{5.3}$$

where, $R_B$ is the volumetric background counting rate (cpm), $V_{sample}$ and $V_B$ are the sample and background air volumes, respectively, $T_{sample}$ and $T_B$ are the sample and background counting times, respectively, $\varepsilon$ is the alpha-particle detection efficiency (dpm/cpm), and $K_B$ is the updated Kusnetz factor (from Table [1]) that corresponds to a delay time determined since end of sampling and using a sum time given by the following equation:

$$Sum \ Time = T_{sample}(1 + \frac{T_{sample}}{T_B}) \tag{5.4}$$

## 7. Conclusions

In this work, sampling- and counting-time corrections are applied to the Kusnetz Method, which is widely used for radon-progeny concentration measurements in air. A table of the updated factors for different equilibrium scenarios is provided in this work for a number of sampling and counting intervals. This study concludes that there is no influence of the equilibrium condition on the factors that correspond to a delay time within the window defined in the Kusnetz Method (i.e. 40-90 minutes after end of sampling the air). Furthermore, ignoring the corrections derived in this work results in 40%, and 17% of underestimation of the working level measurement in air for sampling and counting total times of 40 and 20 minutes, respectively. Finally, a time-corrected self-absorption factor formula and a Kusnetz Method specific LLD formula are provided in this technical report.

## References

[1] U.S. Environmental Protection Agency (EPA). *A citizen's guide to radon: The guide to protecting yourself and your family from radon*, A Citizen's Guide to Radon: The Guide to Protecting Yourself and Your Family from Radon | US EPA, (2016).

[2] D.A. Holaday, D.E. Rushing, R.D. Coleman, P.E. Woolrich, H.L. Kusnetz, and W.F. Bale, *Control of radon and daughters in uranium mines and calculations of biological effects*, (Publication No. 494). U.S. Public Health Service, (1957).

[3] H.L Kusnetz, *Radon Daughters in Mine Atmospheres – A Field Method for Determining Concentrations*, Industrial Hygiene Quarterly, (1956).

[4] W.W. Nazaroff, and A.V. Nero, *Radon and its decay products in indoor air*. John Wiley & Sons, (1988).

[5] T.B. Borak, *A method for prompt determination of working level using a single measurement of gross alpha activity*, Radiation Protection Dosimetry, 19(2), pp. 97-102, (1987)

[6] R.F. Holub, and R.F. Droullard, *Role of Plateout in Radon Daughter Mixture Distributions in Uranium Mine Atmospheres*, Health Phys. 39(5), 761, (1980).

[7] Canadian Nuclear Safety Commission's Regulatory Guide G-4, *Measuring Airborne Radon Progeny at Uranium Mines and Mills*, Regulatory Guide G-4 - Measuring Airborne Radon Progeny at Uranium Mines and Mills (nuclearsafety.gc.ca), (2003).